\documentclass[aip, apl, showpacs, showkeys, epsfig, reprint]{revtex4-1}
\usepackage{graphicx,amsmath}
\usepackage{comment}
\usepackage{xcolor}

\newcommand{\ket}[1]{\mbox{$|{#1}\rangle$}}

\begin{document}

\title{Quantum diamond microscopy with optimized magnetic field sensitivity and sub-ms temporal resolution} 
\author{Sangwon Oh}
\email[]{sangwon.oh@kriss.re.kr}
\affiliation{Quantum Magnetic Imaging Team, Korea Research Institute of Standards and Science, Daejeon 34113, South Korea}
\author{Seong-Joo Lee}
\affiliation{Quantum Magnetic Imaging Team, Korea Research Institute of Standards and Science, Daejeon 34113, South Korea}
\author{Jeong Hyun Shim}
\affiliation{Quantum Magnetic Imaging Team, Korea Research Institute of Standards and Science, Daejeon 34113, South Korea}
\affiliation{Department of Applied Measurement Science, University of Science and Technology, Daejeon 34113, South Korea}
\author{Nam Woong Song}
\affiliation{Quantum Magnetic Imaging Team, Korea Research Institute of Standards and Science, Daejeon 34113, South Korea}
\author{Truong Thi Hien}
\affiliation{Quantum Magnetic Imaging Team, Korea Research Institute of Standards and Science, Daejeon 34113, South Korea}

\date{\today}

\begin{abstract}
Quantum diamond magnetometers using lock-in detection have successfully detected weak bio-magnetic fields from neurons, a live mammalian muscle, and a live mouse heart. This opens up the possibility of quantum diamond magnetometers visualizing microscopic distributions of the bio-magnetic fields. Here, we demonstrate a lock-in-based wide-field quantum diamond microscopy, achieving a mean volume-normalized per pixel sensitivity of 43.9 $\mathrm{nT\mu m^{1.5}/Hz^{0.5}}$. We optimize the sensitivity by implementing a double resonance with hyperfine driving and magnetic field alignment along the $<$001$>$ orientation of the diamond. Additionally, we show that sub-ms temporal resolution ($\sim$ 0.4 ms) can be achieved while keeping the per pixel sensitivity at a few tens of nanotesla per second using quantum diamond microscopy. This lock-in-based diamond quantum microscopy could be a step forward in mapping functional activity in neuronal networks in micrometer spatial resolution.
\end{abstract}


\maketitle

\section{Introduction}
A nitrogen-vacancy (NV) center in diamond is an atomic defect in which magnetic sensitivity is in the range of several microtesla with the nanometer spatial resolution at room temperature ~\cite{tay08,maz08}. The magnetic field sensitivity even reaches below 1 pT/Hz$^{0.5}$ by adopting NV ensembles at the cost of the spatial resolution ~\cite{bar16,fes20,zha21c}. Various techniques such as high-density NV centers ~\cite{bar16}, improved readout fidelity ~\cite{wol15, cle15, dum13}, and flux concentrators~\cite{fes20, zha21c} have been implemented to improve sensitivity. These techniques have contributed to the detection of bio-magnetic fields from neurons, mammalian muscles, and a heart using NV ensembles ~\cite{bar16, web21a, ara22}. However, these measurements exhibit millimeter-scale spatial resolutions, which may limit the visualization of functional activity in neural networks ~\cite{hal12, kar18, kar21}.   
A widefield NV microscope based on the frequency shift of optically detected magnetic resonance (ODMR) can be an alternative method for detecting magnetic field distributions at a micrometer-scale spatial resolution ~\cite{ste10, sim16, gle15, sch21}. Currently, density distribution in graphene, magnetic field arising from an integrated circuit, a 2-D magnet, geological samples, and living cells were imaged at sub- or several-micrometer spatial resolutions ~\cite{tet17, tur20, bro20, gle17, sag13}. However, the widefield NV microscopes based on the ODMR frequency shift may have difficulties in detecting several tens of nanotesla, which corresponds to several hundreds of Hz in the frequency shift. This can be addressed using lock-in-based NV magnetometry ~\cite{sch11}. {A lock-in camera can have a lower noise level than that of conventional CCD(charge-coupled device) or sCMOS(scientific complementary metal oxide semiconductor) cameras when detecting low-frequency signals such as from near dc to 1000 Hz. By modulating the low-frequency signals and then filtering them with a narrow bandwidth, lock-in detection can bypass $1/f$ noise in the signals.}

\begin{figure}[t]
\centering
\includegraphics[width=\columnwidth]{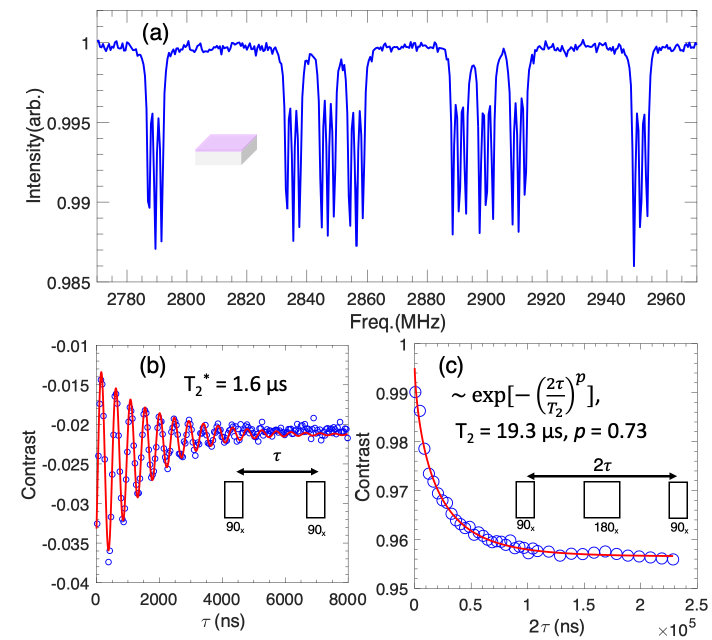}
\caption{ODMR spectrum and decoherence times of the overgrown diamond layer. The ODMR spectrum(a) is {taken when the external magnetic field is not aligned along the $<$111$>$ direction of the NV diamond. Whereas the decoherence times, $\mathrm{T_2^*}$(b) and $\mathrm{T_2}$(c), are measured when the external field is applied along to the $<$111$>$ direction of the NV diamond.} The spectrum and decoherence times support the high quality of the crystal.}
\label{fig1}
\end{figure}

   In this study, we adopt a lock-in camera {and optimize} the magnetic field sensitivity in a wide-field NV microscope ~\cite{woj18a, har21, web22, par22b}. For a simple yet reliable operation, we continuously excite NV diamond using a 532nm laser and a microwave. Multiple hyperfine transitions are simultaneously excited for a higher ODMR contrast, and double resonance is implemented to suppress the influences of temperature drift and strain ~\cite{bar16, fes20, har21, fan13, mam15, shi22}. Additionally, all four NV axes are exploited by aligning an external magnetic field along the $<$001$>$ direction of the NV diamond. Sub-millisecond temporal resolution, 0.4 ms, is demonstrated by detecting a magnetic field from a short-pulsed current at tens of nanotesla per-pixel sensitivity.

\section{Experimental}
A thin nitrogen-doped ([$^{14}$N] $\sim$ 10 ppm) diamond layer ($^{12}$C $>$ 99.99$\%$, 40 $\mu$m thick) is grown by chemical vapor deposition (CVD) on top of an electronic grade diamond plate by Applied Diamond Inc.. The dimensions of the diamond plate are approximately 2 $\times$ 2 $\times$ 0.54 mm$^3$. The diamond is electron irradiated {(1 MeV, 5 $\times$ 10$^{18}$/cm$^2$)}, and annealed in a vacuum at 800$^{\circ}$C for 4 hours and 1000 $^{\circ}$C for 2 hours, sequentially. An ODMR spectrum of the crystal {when the crystal is not aligned along the $<$111$>$ direction of the crystal is shown in Fig.~\ref{fig1}(a). Then we adjust the external magnetic field along the $<$111$>$ direction of the crystal to measure the decoherence times,$\mathrm{T_2^*}$ and $\mathrm{T_2}$. $\mathrm{T_2^*}$ and $\mathrm{T_2}$ }are found to be 1.6 $\mathrm{\mu s}$ and 19.3 $\mathrm{\mu s}$ at B $\approx$ 3 mT, respectively. $\mathrm{T_2^*}$ is found by fitting the data to $C_0$exp[-$(\tau/T_2^*)$]$\sum_{i}$ cos[$\omega_i \tau]$, where $C_0$ is the maximal ODMR contrast, and $\omega_i$ is frequency due to the NV hyperfine splittings, Fig.~\ref{fig1}(b). Similarly, $\mathrm{T_2}$ is obtained by fitting the data to $C_0$exp[-$(2\tau/T_2)^p$], where $p$ is stretched exponential parameter, Fig.~\ref{fig1}(c)~\cite{par22,bau18}.

\begin{figure}[t]
\centering
\includegraphics[width=\columnwidth]{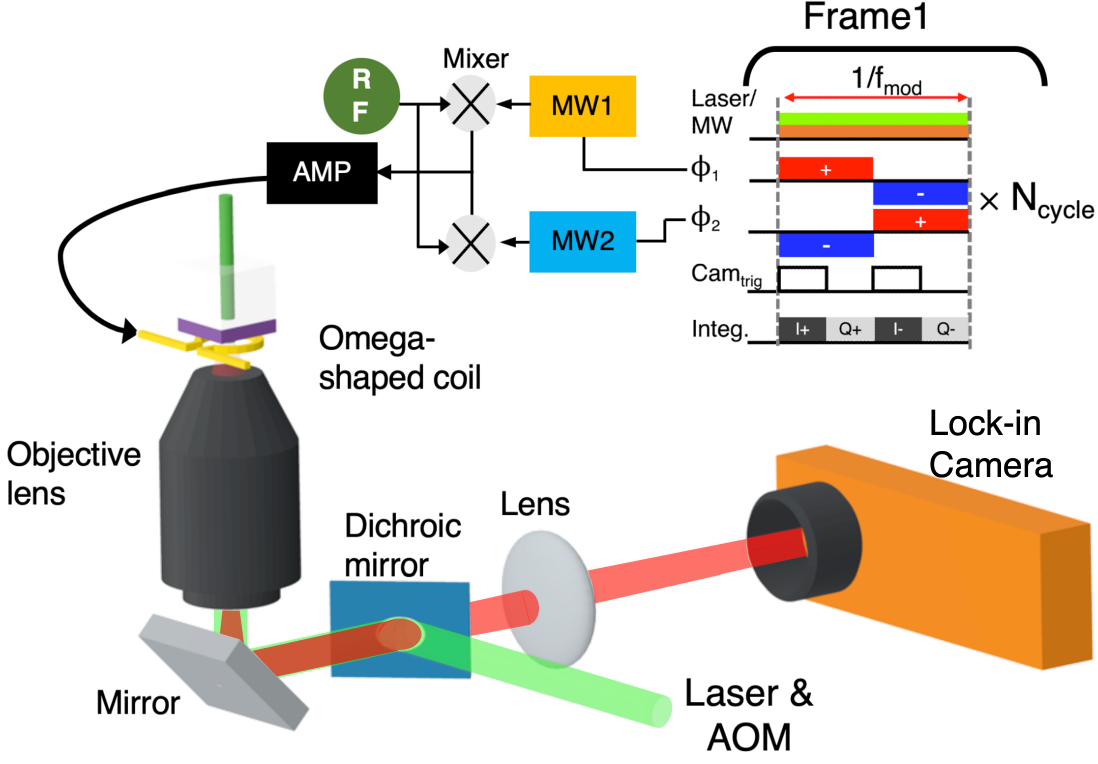}
\caption{(a) Schematic of the widefield microscope and protocol for lock-in camera detection. Frequency modulated (square wave) microwave (MW) is delivered into an omega-shaped coil for lock-in detection. Two MW sources (MW1 and MW2) and one RF (2.16 MHz) generator are mixed and combined for double resonance and hyperfine driving. A continuous-wave (CW) double resonance lock-in protocol is described in the inset. Phases ($\Phi_1$ and $\Phi_2$) of the frequency modulation are synchronized to a camera trigger for signal integration (I$^+$, Q$^+$, I$^-$, and Q$^-$). For magnetic field/temperature detections,  $|\Phi_1 -\Phi_2| $ were kept at $\pi / 0 $, respectively. The trigger, Cam$\mathrm{_{trig}}$, starts the acquisition of the fluorescence, and a single frame of in-phase {(I = I$^+$ - I$^-$)} and quadrature (Q = Q$^+$ - Q$^-$) images are found after repeating$N_{cycle}$ times.}
\label{fig2}
\end{figure}

A schematic of the experimental setup is shown in Fig.~\ref{fig2}. An omega-shaped coil is placed on the overgrown diamond layer and a 532 nm laser (Opus 3W, LaserQuantum) illuminates the diamond. The incident light power and {the beam diameter on the diamond are approximately 200 mW and 46 $\mu m$, respectively. The power density is $\approx$ 0.12 mW/$\mu m^2$.} An objective lens (MPLFN50x, Olympus) is used to collect the red fluorescence from the diamond. Long pass (BLP01-633R, Semrock) and dichroic (LM01-552, Semrock) filters are placed before the lock-in camera (heliCam C3, Heliotis) to separate the 532-nm pump laser. We use two microwave generators (SG394, SRS) and a single RF source (2.16 MHz, WX1282C, Tabor Elec.) for the double resonance with hyperfine driving, {as will be explained below}.  Each SG394 was mixed (ZX05-43MH+, mini-circuits) with WX1282C and then the outputs are combined (ZX10-2-42S+, mini-circuits) for the double resonance with hyperfine driving. The mixed and combined signals are sent to an amplifier (ZHL-16W43-S+, mini-circuits). A switch (ZASWA-2-50DRA+, mini-circuits) is placed before the amplifier to control the delivery of the microwave to the omega-shaped coil. A TTL pulse generator (PB24-100-4k-PCI, spincore) is used to control the switch. 

A frequency modulation (square wave) is selected for the lock-in detection. The modulation depth is 300 kHz and the modulation frequencies (f$_{mod}$) are 2.5 or 10 kHz, depending on the temporal resolution~\cite{ele17}. The phases ($\phi_1, \phi_2$) of the frequency modulation are controlled using an arbitrary waveform generator (AWG, 33522B, Keysight). For magnetic field and temperature detections, their phase differences are maintained at 180$^{\circ}$ and 0$^{\circ}$, respectively. The phases used for imaging magnetic fields are shown in Fig.~\ref{fig2}. An external trigger for the camera (Cam$_{trig}$) comes from the TTL pulse generator and is synchronized to the phases. The trigger internally initiates an integration of the fluorescence signals during four periods, (I$^+$, Q$^+$, I$^-$, and Q$^-$). The in-phase {(I = I$^+$ - I$^-$)} and quadrature (Q = Q$^+$ - Q$^-$) images are automatically calculated by the camera. A single cycle is composed of four periods, and a single frame is a repetition of a single-cycle N$_{cyc}$ times.

\begin{figure}[t]
\centering
\includegraphics[width=0.45 \textwidth]{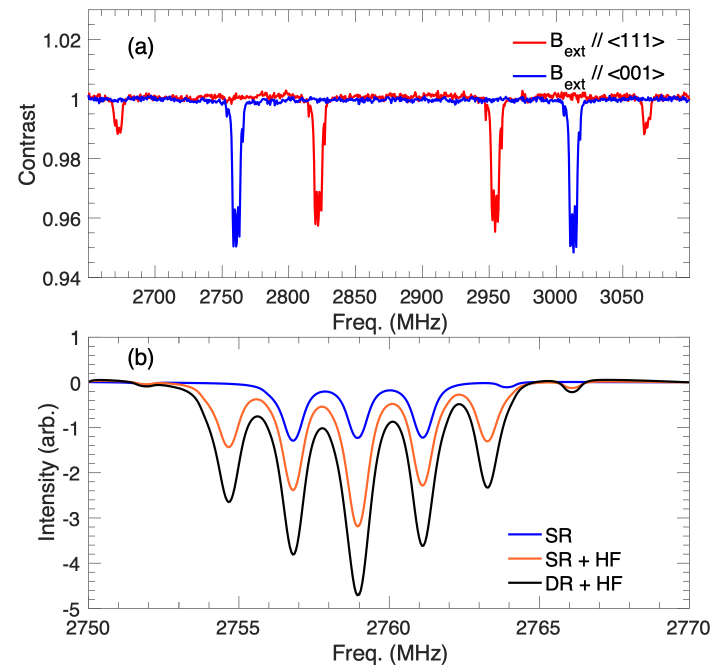}
\caption{ODMR spectra at two aligned cases. (a) ODMR spectra, bias fields aligned along $<$111$>$ and $<$001$>$ of the crystal axes; (b) ODMR of the $<$001$>$ aligned case (SR) compared to the case of hyperfine driving (SR + HF). The contrast of the SR + HF case is enhanced by 2.4 times due to the HF driving. The contrast is further increased by adopting double resonance (DR + HF) with the same phase}
\label{fig3}
\end{figure}

\begin{figure}[t]
\centering
\includegraphics[width=\columnwidth]{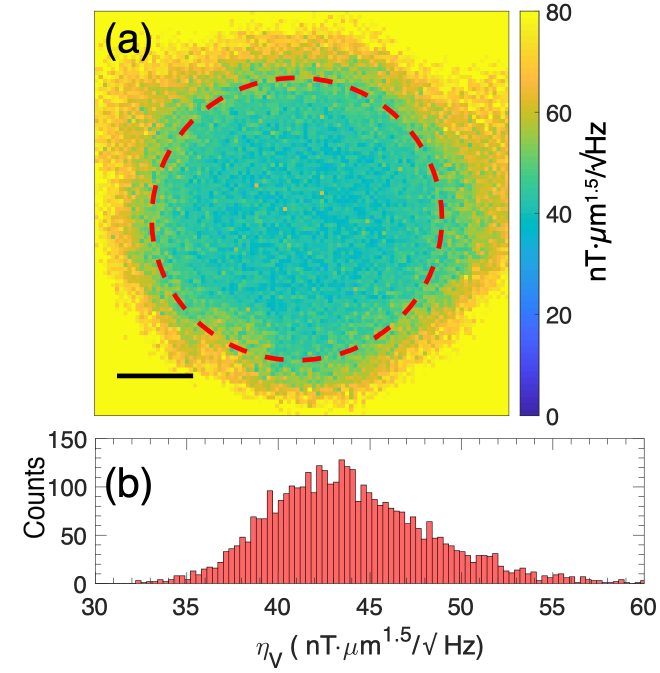}
\caption{Volume-normalized magnetic field sensitivity map. (a) Two-dimensional map of the volume-normalized field sensitivity, $\eta_V$ and the scale bar represents 10 $\mu m$; (b) A histogram of the sensitivity map within the red circled area. Mean volume sensitivity is 43.9 $\mathrm{nT\mu m^{1.5}/Hz^{0.5}}$}
\label{fig4}
\end{figure}

The hyperfine(HF) interaction between the NV and {$^{14}$N} nuclear spin results in a reduced ODMR contrast, which is detrimental to the magnetic field sensitivity. {A single frequency-sweeping ODMR spectrum due to HF interaction can be expressed as 1-$\sum_{p=-1}^{1}\frac{\delta \nu^2}{\delta \nu^2 +4(\omega-(\omega_0 + p \Delta\omega_{HF}))^2}$, where C,$\delta\nu, \omega, \omega_0, \Delta\omega_{HF}$ represent contrast, ODMR linewidth, applied MW frequency, resonant MW frequency, and hyperfine splitting (2.16 MHz in our case), respectively. The central contrast can be enhanced by up to three times if three equally spaced MW frequencies are swept simultaneously, 1-$\sum_{p,q =-1}^{1}\frac{\delta \nu^2}{\delta \nu^2 +4((\omega + q\Delta\omega_{HF})-(\omega_0 + p \Delta\omega_{HF}))^2}$.~\cite{bar16}} In practice, the enhancement is approximately two because of power-broadened hyperfine features.

The negatively charged NV$^-$ has an electronic spin triplet (S = 1) state with a temperature-dependent zero-field splitting, D $\sim$ 2.87 GHz at room temperature, between the \ket{m_s=0} and degenerated \ket{m_s = \pm1}~\cite{doh12}. The degenerated states are splitted by the Zeeman effect as an external magnetic field is applied. The Hamiltonian for the NV$^-$ in a magnetic field {(1 mT $< |B| <$ 10 mT)} can be approximated as: ~\cite{keh19,har21,tet12}

\begin{equation}
\frac{H}{h} = D(T)S_z^2 + \frac{\gamma}{2\pi} B_{NV}S_{z},
\end{equation}

where $z$, S$_z$, and B$_{NV}$, denote the NV symmetry axis, the dimensionless spin-1 operator, the external magnetic field projected along the NV symmetry axis, and $\gamma/2\pi$ is the gyromagnetic ratio (28 GHz/T), respectively. The hyperfine interaction and spin-stress coupling parameters are ignored. The resonance frequencies can be expressed as: ${f_1 = D(t) -\gamma B_{NV}(t)}$ for \ket{m_s=0}$\leftrightarrow$ \ket{m_s=-1} and ${f_2 = D(t) + \gamma B_{NV}(t)}$ for \ket{m_s=0}$\leftrightarrow$\ket{m_s=1} where ${D(t) = D_0 + \Delta D(t)}$ and ${B_{NV}(t) = B_{NV0} + \Delta B_{NV}(t)}$.

\begin{figure}[t]
\centering
\includegraphics[width=\columnwidth]{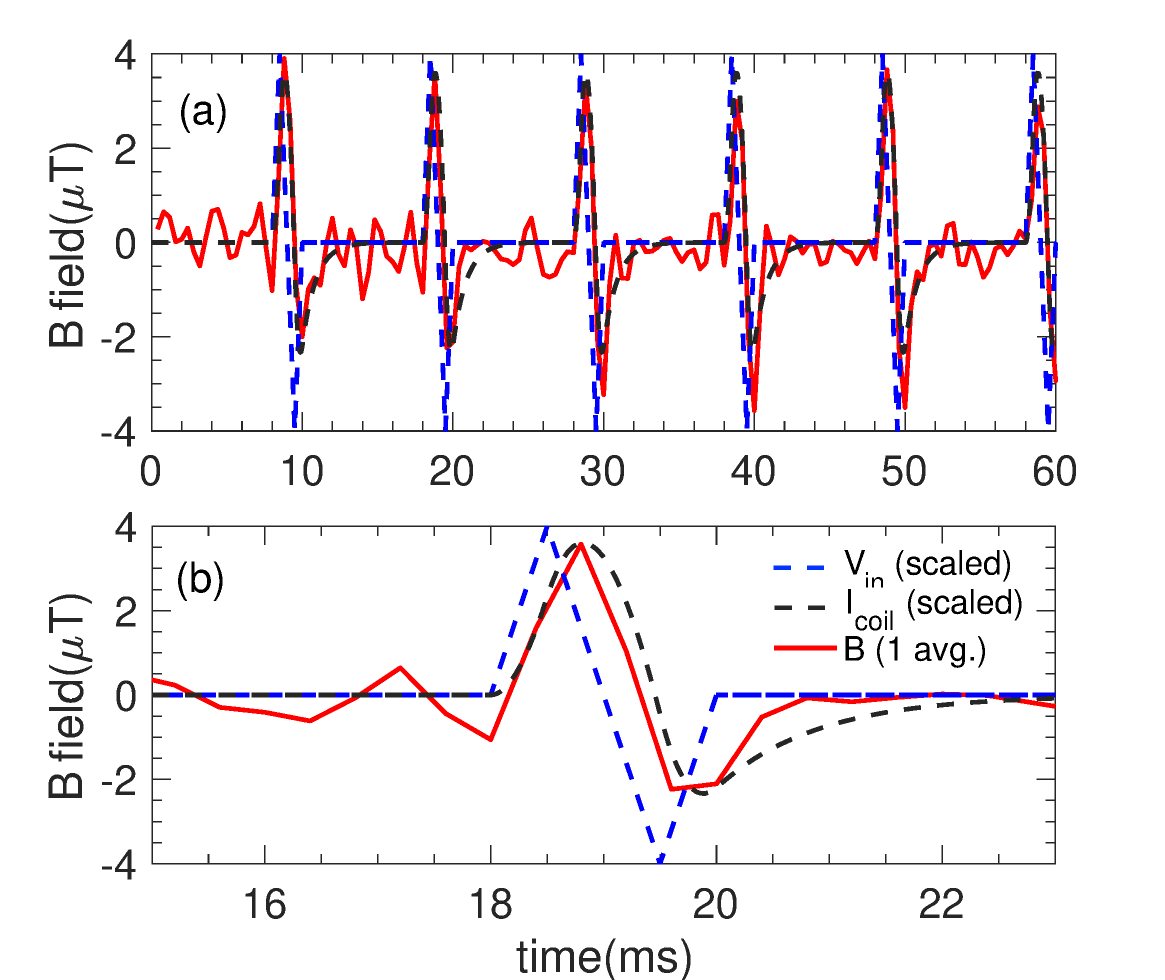}
\caption{Sub-ms temporal resolution. (a) A pulsed (triangular shape) voltage (dashed blue line) is applied at a coil of which inductance and resistance are 1.8 mH and 2 $\Omega$, respectively. An expected current (dashed black line) in the coil is simulated by LTspice. The measured magnetic field (solid red line), parallel to the z-axis of the crystal, is found from a single pixel in the middle of the sensitivity map in Fig.~\ref{fig4} (a) and shown in the upper panel. A detailed view, (b), in the time domain shows that an inductive delay in the current, $\sim$ 0.9 ms, is qualitatively captured by the magnetic field measurement. }
\label{fig5}
\end{figure}

A double resonance (DR) simultaneously drives the resonance frequencies ($f_1$, $f_2$) using two microwave generators (MW1 and MW2) with the modulation frequency ($f_{mod}$) and phases ($\phi_1$, $\phi_2$). The output signals of the lock-in amplifier at $f_1$ and $f_2$ are{ $S_1(t) =\alpha [\Delta D(t) - \gamma \Delta B_{NV}(t)]$ and $S_2(t) =\alpha [\Delta D(t) + \gamma \Delta B_{NV}(t)]$, }respectively, where $\alpha$ is the slope of the lock-in amplifier. If we apply DR with the same phase, i.e. $\phi_1$ = $\phi_2$, then the lock-in signal ($S_{LIA}$) is only sensitive to the temperature, $S_{LIA}$ = 2 $\alpha \Delta$ D(t). Alternatively, it depends only on the magnetic field ($\Delta B_{NV}$) if $ | \phi_1 - \phi_2 |$ = $\pi$ , $S_{LIA}$ = $2 \alpha\Delta B(t)$ ~\cite{fes20, woj18b, shi22, hat21}. Additionally, the sensitivity of the DR method is expected to be enhanced by $\sim$4/3 times compared to the sensitivity of the single-resonance method ~\cite{fes20}. 

\section{Results}
The shot-noise-limited continuous wave (CW) magnetic field sensitivity,$\eta_{CW}$, is given by~\cite{bar16,dre11}:

\begin{equation}
		\eta_{CW} = \frac{4}{3\sqrt3}\frac{h}{g_e\mu_B}\frac{\delta\nu}{C\sqrt R},
\end{equation}

where R is the photon-detection rate, {$\delta\nu$} is the linewidth, and $C$ is the ODMR contrast. To minimize the $\eta_{CW}$, we adopt several methods to obtain a higher ODMR contrast in Eq. (2). The first method is projecting a magnetic field equally along all the NV axes. We compare the ODMR spectra where the external magnetic fields are aligned along the $<$111$>$ and $<$001$>$ directions of the crystal, as shown in Fig.~\ref{fig3}(a). When the magnetic field is along the $<$001$>$ direction of the crystal, the ODMR contrast can be {increased by 2.6 times compared to the ODMR in case of the $<$111$>$ alignment~\cite{bar16,shi22}. The magnetic field sensitivity will be enhanced by 1.5 (= 2.6/$\sqrt3$)times due to the magnetic field projection onto the four NV axes~\cite{fes20}}. Hereafter, we fix the external field along the $<$001$>$ direction. The second method is to simultaneously excite the three HF features (SR +  HF) instead of exciting a single frequency (SR). The ODMR contrast is improved by 2.4 times, compared to that in SR, as shown in Fig.~\ref{fig3}(b). The third method involves applying DR along with HF driving (DR + HF). This further enhances the contrast compared to that in SR + HF, as shown in Fig.~\ref{fig3} (b). DR is essential for minimizing errors in the magnetic field due to temperature drift in the system~\cite{fan13,mam15,shi22}.

The magnetic field sensitivity can be expressed as $\eta = \delta B\sqrt T$, where $\delta B$ is the minimum detectable magnetic field, and T is the measurement duration. The minimum magnetic field is given by the standard deviation of a series of measurements~\cite{har21,par22b} . To determine the minimum magnetic field, a test field is applied along the $z$-axis of the crystal. The test field is found to be 6.8 $\mu$T, and the projected magnetic field along the NV axes is 4 $\mu$T. The frame rate of the camera is 114 Hz (8.8 ms, $f_{mod}$ = 2.5 kHz, 22 cycles), and 110 frames are collected for the estimation.

A two-dimensional map of the volume-normalized magnetic field sensitivity, $\eta_V = \eta \sqrt V$, is shown in Fig. ~\ref{fig4}(a), where the field of view is approximately 46 $\times$ 46 $\mu m^2$, the pixel size is 0.54 $\times$ 0.54 $\mu m^2$, and the sensor volume, V, is 11.7 (0.54 $\times$ 0.54 $\times$ 40) $\mu m^3$. The mean $\eta_V$ within the red circled area is 43.9 $\mathrm{nT\mu m^{1.5}/Hz^{0.5}}$ and the mean per pixel $\eta$ is 12.8 $\mathrm{nT/Hz^{0.5}}$. The histogram of the $\eta_V$ within the area shows a Gaussian-like distribution due to the beam shape, as shown in Fig.~\ref{fig4} (b). Illuminating the NV diamond based on the total internal reflection is expected to result in a larger field of view and a more uniform sensitivity distribution than that in the current scheme~\cite{cle15,har21}.

To demonstrate a sub-millisecond temporal resolution, the frame rate is increased to 2500 Hz, and 200 frames are acquired, where the modulation frequency is set to 10 kHz. The increased modulation frequency decreases the signal-to-noise ratio due to the wider bandwidth. A series of pulsed voltages are applied to a coil with a diameter of 10 cm, an inductance of 1.8 mH, and a resistance of 2 $\Omega$. The voltage pulse has a triangular shape, and its polarity is changed within 2 ms and repeats every 10 ms, as shown in Fig.~\ref{fig5} (b). Because the dimensions of the coil are significantly larger than the field of view (Fig.~\ref{fig4} (a)), the magnetic field produced by the pulsed voltage is uniform.

\begin{figure}[t]
\centering
\includegraphics[width=\columnwidth]{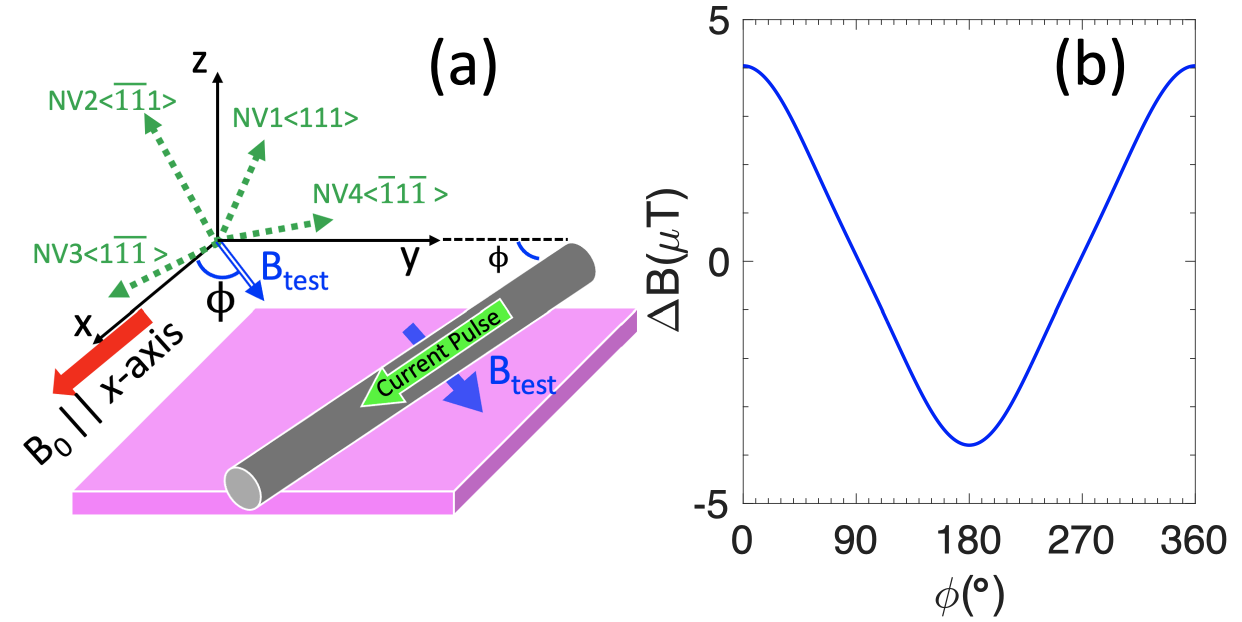}
\caption{{Expected magnetic field response when a test magnetic field is not parallel to the bias field. We calculate sensed magnetic field behavior, $\Delta B(\phi)$, when a test field, $B_{test}$, is not aligned along the bias field. We set the bias field, $B_0$ = 2.6 mT, along the $x$-axis of the NV diamond and rotate $B_{test}$ of 4 $\mu$T around the $z$-axis of the crystal. This represents a case where a current-carrying neuron is placed on an NV-containing diamond plate where the magnetic field from the neuron is parallel to the surface of the diamond.  As Barry $et$ $al.$ claim, an orientation of the neuron could be identified by measuring $\Delta$B if a priori such as the intensity of B$_{test}$ is known~\cite{bar16}. }}
\label{fig6}
\end{figure}

A single acquisition of the magnetic field from the voltage pulse at a single central pixel is shown in Fig.~\ref{fig5}(a). The dashed blue line represents the applied voltage (scaled), and the dashed black line represents the expected current in the coil. The solid red line represents the magnetic field along the $z$-axis of the crystal. Even in a single measurement, we can distinguish $\pm$ 4 $\mu$T magnetic pulse trains from the noise. The expected delay between the current (magnetic field) and the voltage is approximately 0.9 (1.8/ 2) ms. A close-up in the time domain, Fig.~\ref{fig5}(b), indicates that our system can capture the transient behavior with sub-millisecond temporal resolution. The standard deviation of the noise level ($\approx$ 1 $\mu$T) during the acquisition duration (0.4 ms) leads to a per-pixel sensitivity of 20 $\mathrm{nT/Hz^{0.5}}$. These observations support the nanotesla sensitivity {per second} with sub-millisecond temporal resolution.  

\section{Discussion}
In this study, we optimize the volume-normalized magnetic field sensitivity of NV center ensembles using a lock-in camera. The mean per pixel volume-normalized magnetic field sensitivity of 43.9 $\mathrm{nT\mu m^{1.5}/Hz^{0.5}}$ and the sub-ms temporal resolution are obtained at a relatively low optical power density of 0.12 mW/$\mu m^2$. However, we still need to improve the sensitivity to less than 1 $\mathrm{nT\mu m^{1.5}/Hz^{0.5}}$ to visualize neuronal networks~\cite{kar18, kar21}. {The best volume-normalized magnetic field sensitivity obtained by a lock-in camera is 32 $\mathrm{nT\mu m^{1.5}/Hz^{0.5}}$ by Hart $et$ $al.$ and 31 $\mathrm{nT\mu m^{1.5}/Hz^{0.5}}$ using an sCMOS camera by Kazi $et$ $al.$~\cite{har21, kaz21}. The sensitivities are obtained using different NV concentrations.} In this section, we will discuss how we can further improve the magnetic field sensitivity, {spatial resolution, and find the orientation of current-carrying neurons.}

Photonic structures such as diamond nano-pillars will enhance the volume-normalized sensitivity by improving readout fidelity~\cite{mom15, mcc20}. It has been reported that the sensitivity can be improved by more than four times owing to increased photoluminescence and spin coherence time by nano-pillar~\cite{mcc20}. An additional antireflective coating of 600 - 800 nm on the diamond further increases the photoluminescence further ~\cite{web19}.

The inhomogeneous spin dephasing time, $T_2^*$, can be extended by applying decoupling sequences. The dipolar coupling between NV centers and substitutional nitrogen (P1) can be suppressed by driving P1 spins~\cite{bau18, bal19}. Bauch $et$ $al.$ and Balasubramanian $et$ $al.$ reported that $T_2^*$ in a high P1 density increased more than four times using spin-bath driving~\cite{bau18, bal19}. Moreover, Balasubramanian $et$ $al.$ decoupled NV-NV interaction by adopting WAHUHA sequence and additionally extended $T_2^*$ by ten times~\cite{bal19}. These methods, combined with our technique, could reduce the volume-normalized sensitivity to less than 1 $\mathrm{nT\mu m^{1.5}/Hz^{0.5}}$, which is an essential tool for understanding neuronal connectivity~\cite{hal12, kar18, kar21}.

Illuminating the NV layer uniformly using total internal reflection geometry improves the magnetic sensitivity distribution and increases the field of view up to the millimeter scale~\cite{cle15, tur20}. This can be utilized to detect magnetic fields from an integrated circuit and 3-D current distribution in a multi-layer printed circuit board~\cite{tur20, oli21}. Combined with the sub-millisecond temporal resolution, the wide field of view could contribute to imaging transient events, which could be missed by scanning-based systems such as giant-magneto resistive (GMR) or superconducting quantum interference device (SQUID)-based current mapping equipment ~\cite{tur20, gau14}.

{Since our NV thickness is approximately 40 $\mu$m, the spatial resolution is more than the thickness. However, it could be useful for a millimeter-scale wide field of view system with a few tens of micrometer pixel size. The per-pixel magnetic field sensitivity could be about 200 pT/Hz$^{0.5}$ if the pixel is 40 $\mu$m. }

{We calculate how the sensed magnetic field, $\Delta B$, changes when a test field, $B_{test}$, is misaligned from the bias magnetic field, $B_0$. To match our experimental conditions, we set $B_0$ = 2.6 mT and $B_{test}$ = 4 $\mu$T, respectively. We assume that $B_0$ is parallel to the $x$-axis of the crystal and $B_{test}$ is rotated by $\phi$ from the $x$-axis and parallel to the $xy$ plane, as shown in Fig.~\ref{fig6}(a). This represents a case where a current-carrying neuron is placed on an NV-containing diamond plate. The magnetic field from the neuron, B$_{test}$, is parallel to the surface of the diamond plate and encircles the $z$-axis of the crystal depending on the orientation of the neuron. We find that the magnetic field, $\Delta B(\phi)$, sensed by the NV diamond changes by a factor of $\mathrm{cos(\phi)}$ by solving Eq. 1 as shown in Fig.~\ref{fig6}(b). As Barry $et$ $al.$ claim, an orientation of the neuron could be identified by measuring $\Delta$B if a priori such as the intensity of B$_{test}$ is known~\cite{bar16}. This could give a way to image neuronal connectivity using NV centers with further optimization in the sensitivity.}

\section{Conclusion}

In conclusion, we have obtained a mean per pixel volume-normalized magnetic sensitivity of 43.9 $\mathrm{nT\mu m^{1.5}/Hz^{0.5}}$ and a sub-ms temporal resolution using NV center ensembles and a lock-in camera. The HF driving, DR, and exploitation of the four NV axes are adopted with CW lock-in detection to reach the sensitivity. These methods could be a step forward for visualizing microscopic distributions of sub-nanotesla changes due to neuronal currents in real-time, as well as defects in a packaged battery~\cite{zha21x}.

\section*{Declarations}

\subsection*{Acknowledgments}
The authors thank Kiwoong Kim for valuable discussions and Heloitis AG for experimental assistance in implementing the camera.
\subsection*{Funding}
This research was supported by Institute of Information $\&$ communications Technology Planning $\&$ Evaluation (IITP) grants funded by the Korea government (MSIT) (No.2019-000296, No.2021-0-00076) and a grant (GP2021-0010) from Korea Research Institute of Standards and Science.

\subsection*{Availability of data and materials}
The data that support the findings of this study are available from the corresponding author upon reasonable request.

%

\end{document}